\documentclass[a4paper]{jpconf}

\usepackage{latexsym}
\usepackage{amsmath}
\usepackage{amssymb}
\usepackage{amsfonts}
\usepackage{bm}

\usepackage{graphicx}
\usepackage{textcomp}

\begin{document}

\title{\hspace*{-0.25cm} Effective field theory investigations of the XYZ 
puzzle}

\author{Jorge Segovia}
\address{Physik-Department, Technische Universit\"at M\"unchen, \\ 
James-Franck-Str. 1, 85748 Garching, Germany}

\ead{jorge.segovia@tum.de}

\begin{abstract}
Quantum Chromodynamics, the theory of strong interactions, predicts several 
types of bound states. Among them are mesons ($q\bar{q}$) and baryons ($qqq$), 
which have been the only states observed in experiments for years. However, in 
the last decade, many states that do not fit this picture have been observed at 
$B$-factories (BaBar, Belle and CLEO), at $\tau$-charm facilities (CLEO-c, 
BESIII) and also at proton-proton colliders (CDF, D0, LHCb, ATLAS, CMS). There 
is growing evidence that at least some of the new charmonium- and 
bottomonium-like states, the so-called XYZ mesons, are new forms of matter such 
as quark-gluon hybrids, mesonic molecules or different arrangements of 
tetraquarks, pentaquarks... Effective Field Theories (EFTs) have been 
constructed for heavy-quark-antiquark bound states, but a general study of the 
XYZ mesons within the same framework has not yet been done. The scope of this 
conference proceedings is to discuss the possibilities we have in developing 
novel EFTs that, characterizing the conventional quarkonium states, facilitate 
also the systematic and model-independent description of the new exotic  
matter, in particular, the hybrid mesons.
\end{abstract}


\section{Introduction}

One of the basic properties of Quantum Chromodynamics (QCD) is its spectrum: 
the list of particles that are stable or at least sufficiently long-lived to be 
observed as resonances. The elementary constituents in QCD are quarks $(q)$, 
antiquarks $(\bar{q})$, and gluons $(g)$, and QCD requires that they be 
confined into color-singlet clusters called hadrons. The most stable hadrons 
are the clusters predicted by the quark model~\cite{GellMann:1964nj, 
Zweig:1964CERN}, mesons $(q\bar{q})$ and baryons $(qqq)$, which have been the 
only states observed in experiments for years~\cite{Agashe:2014kda}).

A decade ago, the Belle Collaboration discovered an unexpected enhancement at 
$3872\,{\rm MeV}$ in the $\pi^{+}\pi^{-}J/\psi$ invariant mass spectrum while 
studying the reaction $B^{+}\to K^{+}\pi^{+}\pi^{-}J/\psi$~\cite{Choi:2003ue}. 
The $X(3872)$ state was later studied by CDF, D0, and BaBar collaborations 
confirming that its quantum numbers, mass and decay patterns make it an unlikely 
conventional charmonium candidate. Therefore, the simple picture that had been 
so successful for $30$ years was challenged leading to an explosion of related 
experimental activity. The number of new collectively denoted XYZ states has 
increased dramatically. Almost two dozen charmonium- and bottomonium-like XYZ 
states have forced an end to the era when heavy quarkonium was considered as a 
relatively well established heavy quark-antiquark bound system (see, e.g., 
reviews~\cite{Brambilla:2010cs, Brambilla:2014jmp} for more details).

The ultimate goal of theory is to describe the properties of the XYZ states 
from QCD's first principles. However, since this task is quite challenging, a 
more modest goal to start with is the development of QCD motivated 
phenomenological models that specify the colored constituents, how they are 
clustered and the forces between them. For this reason, at the same time that 
many experimental measurements on XYZ physics were performed, theorists proposed 
different kind of exotic hadrons to classify them:
\begin{enumerate}
\item[1)] {\it A Quarkonium hybrid,} which has constituents $Q\bar{Q}g$, where 
$g$ is a constituent gluon. The $Q\bar{Q}$-pair is in a color-octet state 
because the color charge of a gluon is octet.
\item[2)] {\it A Tetraquark,} which consists on one heavy quark $(Q)$, one 
heavy antiquark $(\bar{Q})$, one light quark $(q)$ and one light antiquark 
$(\bar{q})$ whose different arrangements allowed by QCD give:
\begin{enumerate}
\item[2.1)] {\it A meson molecule}~\cite{Tornqvist:1993ng}, which consists of 
color-singlet $(Q\bar{q})_{1}$ and $(\bar{Q}q)_{1}$ mesons bound by 
pseudoscalar-meson-exchange interactions.
\item[2.2)] {\it A diquarkonium}~\cite{Drenska:2010kg}, which consists of a 
color-antitriplet $(Qq)_{\bar{3}}$ diquark and a color-triplet 
$(\bar{Q}\bar{q})_{3}$ anti-diquark bound by gluon exchanges.
\item[2.3)] {\it A hadro-quarkonium}~\cite{Dubynskiy:2008mq}, which consists 
of a color-singlet $(q\bar{q})_{1}$-pair bound through color van der Waals 
forces to a compact color-singlet $(Q\bar{Q})_{1}$-pair.
\item[2.4)] {\it A quarkonium adjoint meson or Born-Oppenheimer 
tetraquark}~\cite{Braaten:2013boa}, which consists of two bounded 
color-octets, $(Q\bar{Q})_{8}$ and $(q\bar{q})_{8}$, with dynamics similar to 
that of hybrids.
\item[2.5)] {\it A compact tetraquark}~\cite{Vijande:2007rf}, in which the 
constituents do not form any distinguishable cluster but form as a whole a 
color singlet state: $(Q \bar Q q \bar q)_1$.
\end{enumerate}
\item[3)] {\it A Pentaquark and so on}, which are the various multiquark 
configurations allowed by QCD that one can imagine and have not been already 
cited.
\end{enumerate}

Thus far, the phenomenological descriptions specified above have produced 
compelling explanations for some of the XYZ states but not for all of them 
in an unified way. It would be desirable to have a single theoretical framework 
based firmly on QCD that describes all the XYZ mesons.


\section{Nonrelativistic effective field theories for conventionals}

Heavy quarkonium systems are characterized by their nonrelativistic nature, i.e.
the heavy quark bound-state velocity, $v$, satisfies $v\ll1$. This is reasonably
fulfilled in bottomonium ($v^{2} \sim 0.1$) and to a certain extent in 
charmonium ($v^{2}\sim 0.3$). Moreover, at least, three widely separated scales
appear: the heavy quark mass $m$ (hard scale), the relative momentum of the 
bound state $p\sim mv$ (soft scale) and the binding energy $E\sim mv^{2}$ 
(ultrasoft scale). With $v\ll1$, the following hierarchy of scales
\begin{equation}
m \gg p \sim 1/r \sim mv  \gg E \sim mv^2\,
\end{equation}
is satisfied and this allows for a description in terms of EFTs of physical 
processes taking place at one of the lower scales. The integration out of modes 
associated with high-energy scales is performed as part of a matching procedure 
that enforces the equivalence between QCD and the EFT at a given order of the 
expansion in $v$. The final result is a factorization at the Lagrangian level 
between the high-energy modes, which are encoded in the matching coefficients, 
and the low-energy contributions carried by the dynamical degrees of freedom.

\subsection{Physics at the scale $m$: NRQCD}

The suitable EFT to describe heavy quarkonium annihilation and production, 
which take place at the scale $m$, is Nonrelativistic QCD 
(NRQCD)~\cite{Caswell:1985ui, Bodwin:1994jh}. It follows from QCD by 
integrating out the high energy modes of order $m$. As a consequence, the 
effective Lagrangian is organized as an expansion in $1/m$ and $\alpha_{s}(m)$: 
\begin{equation}
{\cal L}_{\rm NRQCD}  = \sum_n \frac{c_n(\alpha_{s}(m),\mu)}{m^{n}} 
\times  {\cal O}_n(\mu,mv,mv^2,\ldots) \,,
\label{eq:L_NRQCD}
\end{equation}
where $c_n$ are the Wilson coefficients that encode the contributions from the 
scale $m$, $\mu$ is the NRQCD factorization scale, and ${\cal O}_n$ are the 
low-energy operators constructed out of two or four heavy-quark/antiquark fields 
plus gluons. The matrix elements of ${\cal O}_n$ depend on the scales $\mu$, 
$mv$, $mv^2$ and $\Lambda_{\rm QCD}$. Thus, the operators are counted in powers 
of $v$. The imaginary part of the coefficients of the four-fermion operators 
contains the information on heavy quarkonium annihilation and production.

The NRQCD heavy quarkonium Fock state is given by a series of terms, where the 
leading term is a $Q\bar{Q}$ in a color-singlet state, and the first 
correction, suppressed in $v$, comes from a $Q\bar{Q}$ in an octet state 
plus a gluon. Higher-order terms are subleading in increasing powers of $v$.

\subsection{Physics at the scale $mv$: pNRQCD}

The suitable EFT to describe heavy quarkonium formation, which takes place at 
the scale $mv$, is potential NRQCD (pNRQCD)~\cite{Pineda:1997bj, 
Brambilla:1999xf}. It follows from NRQCD by integrating out the modes of order 
$mv \sim 1/r$. The specific construction details of pNRQCD are slightly 
different depending upon the relative size between the soft, $mv$, and the 
confinement, $\Lambda_{\rm QCD}$, scales. Two main situations are 
distinguished. When $mv\gg \Lambda_{\rm QCD}$, we speak about weakly-coupled 
pNRQCD because the matching of NRQCD to pNRQCD may be performed in perturbation 
theory. When $mv\sim \Lambda_{\rm QCD}$, we speak about strongly-coupled pNRQCD 
because the matching of NRQCD to pNRQCD is not possible in perturbation theory. 

When the soft scale is perturbative, the effective pNRQCD Lagrangian is 
organized as an expansion in $1/m$ and $\alpha_{s}(m)$, inherited from NRQCD, 
but also as a multipole expansion in $r$:
\begin{equation}
{\cal L}_{\rm pNRQCD} = \int d^3r\,  \sum_n \sum_k 
\frac{c_n(\alpha_{s}(m),\mu)}{m^{n}} \, V_{n,k}(r,\mu^\prime,\mu) \, r^{k}  
\times {\cal O}_k(\mu^\prime,mv^2,\ldots) \,,
\label{eq:L_pNRQCD}
\end{equation}
where ${\cal O}_k$ are the operators of pNRQCD that depend on the scales 
$\mu^\prime$, $mv^2$ and $\Lambda_{\rm QCD}$; the pNRQCD factorization scale 
is $\mu^\prime$, and $V_{n,k}$ are the Wilson coefficients that encode the 
contributions from the scale $r$.

The degrees of freedom that make up the operators ${\cal O}_k$ are 
$Q\bar{Q}$-pairs in a color-singlet or a color-octet state, and ultrasoft 
gluons. As one can see in Eq.~(\ref{eq:L_pNRQCD}), the Wilson coefficients 
$V_{n,k}$ depend on the relative distance $r$ between the quark and the 
antiquark. They are potential-like terms when $k=0$ and are the $1/m^n$ 
potentials that enter in a Schr\"odinger-type equation. Non-potential 
interactions, $V_{n,k\neq0}$, account for singlet to octet transitions via 
ultrasoft gluons and provide loop corrections to the leading potential picture. 
They are typically related to nonperturbative effects.


\section{Non-relativistic effective field theories for exotics}

The above EFTs have proven to be very successful for the description of 
conventional heavy quarkonium states which are below the open-flavour threshold 
($D\bar{D}$ ($B\bar{B}$) in the charmonium (bottomonium) sector). However, 
no EFT description has yet been constructed for those states which are close to 
or above threshold because the dynamical situation changes 
drastically~\cite{Vairo:2006pc, Brambilla:2008zz}.

The threshold regions are actually the most interesting ones because many of 
the charmonium- and bottomonium-like XYZ states are located in such regions. 
Ab-initio lattice-regularized QCD computations are also in trouble when dealing 
with threshold regions. Moreover, calculations of excited states using this 
technique have been only recently pioneered and the full treatment of 
bottomonium on the lattice seems to be tricky. This all together explains why 
many of our expectations for the XYZ states still rely on phenomenological 
models.

Below we examine how things change when studying one particular case of the XYZ 
particles: the quark-gluon hybrids, and elucidate the future steps to be done 
towards an EFT description of them. It is important to keep in mind that the 
XYZ states belong to heavy quark sectors and thus the heavy quark mass is still 
an appropriate parameter from which begin a nonrelativistic expansion.

\subsection{An example: hybrid states}

Consider the case in which a system is made by a heavy quark, a heavy antiquark 
and gluonic excitations aiming to describe heavy quarkonium hybrids.
In the static limit, at and above the $\Lambda_{\rm QCD}$ threshold, a tower of 
hybrid static energies (i.e. of gluonic excitations) must be considered on top 
of the $Q\bar{Q}$ static singlet energy~\cite{Horn:1977rq, Hasenfratz:1980jv}. 
The spectrum has been thoroughly studied in lattice NRQCD~\cite{Juge:2002br} 
through the logarithm of large time generalized static Wilson loops divided by 
the interaction time:
\begin{equation}
E_{n}^{(0)}(r) = \lim_{T\to \infty} \frac{i}{T} \log \left\langle\right.\!\! 
X_{n}, T/2 | X_{n}, -T/2 \!\!\left.\right\rangle \,, 
\end{equation}
where the NRQCD initial and final states can be constructed as follows
\begin{equation}
|X_n\rangle = 
\chi(\bm{x}_2) \phi(\bm{x}_2,\bm{R}) T^aP_n^a(\bm{R}) \phi(\bm{R},\bm{x_1}) 
\psi^\dagger(\bm{x}_1) |\mathrm{vac}\rangle\,.
\end{equation}
with $\phi(\bm{x}_2,\bm{x}_1)$ a Wilson line from $\bm{x}_1$ to $\bm{x}_2$, and 
$P_n$ is some gluonic operator that generates the desired quantum numbers 
$n$ to calculate the static energy. A list of possible operators $P_n$ can be 
found, e.g., in Refs.~\cite{Brambilla:1999xf, Bali:2003jq}.

Berwein {\it et al.}, based on Refs.~\cite{Brambilla:1999xf, Bali:2003jq, 
Brambilla:2000gk, Pineda:2000sz}, have set in Ref.~\cite{Berwein:2015vca} the 
first steps towards a pNRQCD characterization of the gluonic energies and thus 
of the hybrid potential. The matching between NRQCD and pNRQCD allows to 
establish that the spectrum of the hybrid static energies is described at short 
distances and in the leading multipole expansion of pNRQCD by the octet 
potential plus a mass scale called gluelump mass:
\begin{equation}
E_{n}^{(0)}(r) = \lim_{T\to \infty} \frac{i}{T} \log \left\langle\right.\!\! 
X_{n}, T/2 | X_{n}, -T/2 \!\!\left.\right\rangle = V_{o}(r) + \Lambda_{H} + 
{\cal O}(r^{2}) \,.
\label{eq:pNRQCD_En}
\end{equation}

Equation~(\ref{eq:pNRQCD_En}) can be systematically improved by calculating 
higher orders in the multipole expansion. In particular, one can look at how 
the symmetry group of the hybrid system at short distances $(O(3)\otimes C)$ is 
softly broken to the symmetry group that characterized the gluonic static 
energies in NRQCD ($D_{\infty h})$. We know that the leading correction coming 
from the multipole expansion is at ${\cal O}(r^2)$ and can be calculated in 
pNRQCD in terms of nonperturbative correlators to be eventually evaluated on the 
lattice or in QCD vacuum models. This correction has been treated 
phenomenologically in Ref.~\cite{Berwein:2015vca} and its computation will be 
one of our future goals.

Another of our goals will be going beyond the static limit and compute 
$1/m^{n}$ terms with $n=1,\,2,\,\ldots$ For instance, the NRQCD Hamiltonian  
for the one-quark-one-antiquark sector of the Fock space reads
\begin{align}
H_\mathrm{NRQCD} &= H^{(0)} + \frac{1}{m}H^{(1)} + \ldots\,, \label{HH} \\
H^{(0)} &= \int d^3x\, \frac{1}{2}\left( \bm{E}^a\cdot\bm{E}^a 
+\bm{B}^a\cdot\bm{B}^a \right)-\sum_{j=1}^{n_f} \int d^3\bm{x}\, \bar{q}_j \, i 
\bm{D}\cdot \bm{\gamma} \, q_j \,,\label{H0}\\
H^{(1)} &= - \frac{1}{2} \int d^3x\, \psi^\dagger \left( \bm{D}^2 + g c_F\, 
\bm{\sigma} \cdot \bm{B}\right) \psi + \frac{1}{2}\int d^3x\, \chi^\dagger 
\left(\bm{D}^2+ g c_F\, \bm{\sigma} \cdot \bm{B} \right) \chi\,,\label{H01}
\end{align}
where we have shown only terms up to order $1/m$ in the quark mass expansion, 
$\psi$ ($\chi^{\dagger}$) is the Pauli spinor field that annihilates the heavy 
quark (antiquark), $q_j$ is a massless quark of flavor~$j$, and $c_{F}$ is a 
matching coefficient. We are not considering the light quarks as external 
dynamical sources but they can still appear in the form of sea quarks.

A partial computation of the $1/m$ contributions has been performed 
in Ref.~\cite{Berwein:2015vca}: the kinetic part of $H^{(1)}$ was included 
whereas its spin dependent term was ignored. This has produced spin-multiplets 
which compare nicely with lattice-regularized QCD results (see, for instance, 
Fig.~5 of Ref.~\cite{Berwein:2015vca}), providing the necessary support for 
continuing in this direction. It would be valuable to break the spin degeneracy 
and give a more detailed structure to the hybrid multiplets. 

Finally, let us state here that our long term goal is to introduce an EFT 
description of heavy quarkonium hybrids without using the multipole expansion. 
At large distances, the dynamics of the system is nonperturbative but still 
reduces to a quantum mechanical problem with a number of potentials organized in 
powers of $1/m$~\cite{Brambilla:2004jw}. This would entail the definition of 
appropriate generalized Wilson loops that encode the dynamics of the 
nonperturbative matrix elements and obtaining in strongly-coupled pNRQCD the 
dynamical equations that couple them.


\section{Conclusions}

The search of exotic matter is a topic that has fascinated all generations of 
nuclear and particle physicists since the establishment of QCD as the theory of 
the strong interaction. In this respect, Europe is situated in a privileged 
position for the next decade(s) with two mayor experiments scheduled: LHCb@LHC 
and PANDA@FAIR. In support of the experimental effort, Europeans should also 
play a leading role in the theoretical counterpart. The work presented here 
aims to develop a state-of-the-art theoretical tool able to turn the 
description of exotic matter from a qualitative perspective into a quantitative 
one.


\ack
I would like to thank first Nora Brambilla and Antonio Vairo for insightful 
comments and careful reading of this manuscript;
and second to some of my colleagues who attended the FAIRNESS2016 Workshop, M. 
Albaladejo, M. Berwein, S. Hwang, Z. Meisel, R. Navarro, C. QI, U. Tamponi and 
J. Tarr\'us-Castell\`a, for enjoyable physics discussions.
I also want to express my gratitude to the organizers of the FAIRNESS2016 
Workshop for enabling my participation, which proved very rewarding.
I acknowledge financial support from the Alexander von Humboldt Foundation.


\section*{References}


\bibliographystyle{iopart-num}
\bibliography{JPCS_FAIRNESS2016_JorgeSegovia}

\end{document}